\documentclass[preprint,showpacs,preprintnumbers,amssymb,aps]{revtex4}
\usepackage{amsfonts,color,graphicx,epsfig,amsmath,multirow,slashed}

\begin{document}

\title{Next-to-leading-order QCD corrections to the decay of $Z$ boson into $\chi_c(\chi_b)$}
\author{Zhan Sun$^1$}
\email{zhansun@cqu.edu.cn}
\author{Hong-Fei Zhang$^2$}
\email{hfzhang@ihep.ac.cn}
\affiliation{
\footnotesize
$^{1}$ Department of Physics, Guizhou Minzu University, Guiyang 550025, People's Republic of China. \\
$^{2}$ College of Big Data Statistics, Guizhou University of Finance and Economics, Guiyang, 550025, China.
}

\date{\today}

\begin{abstract}
Based on the framework of nonrelativistic quantum chromodynamics, we carry out next-to-leading-order (NLO) QCD corrections to the decay of $Z$ boson into $\chi_c$ and $\chi_b$, respectively. The branching ratio of $Z \to \chi_{c}(\chi_b)+X$ is about $10^{-5}(10^{-6})$. For the color-singlet (CS) $^3P_J^{[1]}$ state, the heavy quark-antiquark pair associated process serves as the leading role. However the process of $Z \to Q\bar{Q}[^3P_J^{[1]}]+g+g$ can also provide non-negligible contributions, especially for the $\chi_b$ cases. In the case of the color-octet (CO) $^3S_1^{[8]}$ state, the single-gluon-fragmentation diagrams that first appear at the NLO level can significantly enhance the leading-order results. Consequently the CO contributions account for a large proportion of the total decay widths. Moreover, including the CO contributions will thoroughly change the CS predictions on the ratios of $\Gamma_{\chi_{c1}}/\Gamma_{\chi_{c0}}$, $\Gamma_{\chi_{c2}}/\Gamma_{\chi_{c0}}$, $\Gamma_{\chi_{b1}}/\Gamma_{\chi_{b0}}$, and $\Gamma_{\chi_{b2}}/\Gamma_{\chi_{b0}}$, which can be regarded as an outstanding probe to distinguish between the CO and CS mechanism. Summing over all the feeddown contributions from $\chi_{c}$ and $\chi_b$, respectively, we find $\Gamma(Z \to J/\psi+X)|_{\chi_c-\textrm{feeddown}}=(0.28 - 2.4) \times 10^{-5}$ and $\Gamma(Z \to \Upsilon(1S)+X)|_{\chi_b-\textrm{feeddown}}=(0.15 - 0.49) \times 10^{-6}$.

\end{abstract}

\maketitle

\section{Introduction}
As one of the most successful theories describing the production of heavy quarkonium, nonrelativistic quantum chromodynamics (NRQCD) \cite{Bodwin:1994jh} has proved its validity in many processes \cite{Braaten:1994vv,Cho:1995ce,Cho:1995vh,Han:2014jya,Zhang:2014ybe,Gong:2013qka,Feng:2015wka,Han:2014kxa,Wang:2012is,Butenschoen:2009zy,Sun:2017nly,Sun:2017wxk}. Despite these successes, NRQCD still faces some challenges. For example the NRQCD predictions significantly overshoot the measured total cross section of $e^+e^- \to J/\psi+X_{\textrm{non}-c\bar{c}}$ released from the $BABAR$ and Belle collaborations \cite{Zhang:2009ym}. In addition, the polarization puzzle of the hadroproduced $J/\psi$ ($\psi(2S)$) is still under debate \cite{Butenschoen:2012px,Chao:2012iv,Gong:2012ug}. One key factor responsible for these problems is that there are three long distance matrix elements (LDMEs) to be determined, which will bring about difficulties in drawing a definite conclusion.

In comparison with $J/\psi$, $\chi_c$ has its own advantages. First, within the NRQCD framework, in the expansion of $v$ (the typical relative velocity of quark and antiquark in quarkonium) we have
\begin{eqnarray}
|\chi_{QJ}\rangle=\mathcal O(1)|Q\bar{Q}[^3P_{J}^{[1]}]\rangle+\mathcal O(v)|Q\bar{Q}[^3S_{1}^{[8]}]g\rangle+...~.
\end{eqnarray}
$^3S_1^{[8]}$ is the unique color-octet (CO) state involved at the leading-order (LO) accuracy in $v$. From this point of view, $\chi_c$ is more ``clean" comparing to $J/\psi$. In the second place, since the branching ratio of $\chi_c \to J/\psi+\gamma$ is sizeable, the $\chi_c$ feeddown may have a significant effect on the yield and/or polarization of $J/\psi$. For instance including the $\chi_c$ feeddown will obviously make the polarization trend of the hadroproduced $J/\psi$ more transverse. On the experiment side, $\chi_c$ can be easily detected by hunting the ideal decay process, $\chi_c \to J/\psi \to \mu^+\mu^-$. In conclusion, $\chi_c$ is beneficial for studying heavy quarkonium, deserving a separate investigation.

In the past few years, there have been a number of literatures concerning the studies of the $\chi_c$ and $\chi_b$ productions \cite{Cho:1995vh,Cho:1995ce,Chen:2014ahh,Braaten:1999qk,Sharma:2012dy,Shao:2014fca,Li:2011yc,Feng:2015wka,Han:2014jya}. Ma $et$ $al$. \cite{Ma:2010vd} for the first time accomplished the next-to-leading-order (NLO) QCD corrections to the $\chi_c$ hadroproductions. Later on Zhang $et$ $al$. \cite{Jia:2014jfa} carried out a global analysis of the copious experimental data on the $\chi_c$ hadroproduction and pointed out that almost all the existing measurements can be reproduced by the NLO predictions based on NRQCD. To further check the validity and universality of the $\chi_c$ related LDMEs, it is indispensable to utilize them in other processes.

Considering that copious $Z$ boson events can be produced at LHC, the axial vector part of the $Z$-vertex allows for a wider variety of processes, and the relative large mass of $Z$ boson can make the perturbative calculations more reliable, we will for the first time perform a systematic study on the decay of $Z$ boson into $\chi_c$ within the framework of NRQCD. Due to the larger mass of the $b\bar{b}$ mesons, the typical coupling constant and relative velocity of bottomonium are smaller than those of charmonium, subsequently leading to better convergent results over the expansion in $\alpha_s$ and $v^2$ than the charmonium cases. Thus, in this article, the $\chi_b$ productions via $Z$ boson decay will also be systematically investigated.

The rest of the paper is organized as follows: In Sec. II, we give a description on the calculation formalism. In Sec. III, the phenomenological results and discussions are presented. Section IV is reserved as a summary.

\section{Calculation Formalism}

\begin{figure*}
\includegraphics[width=0.95\textwidth]{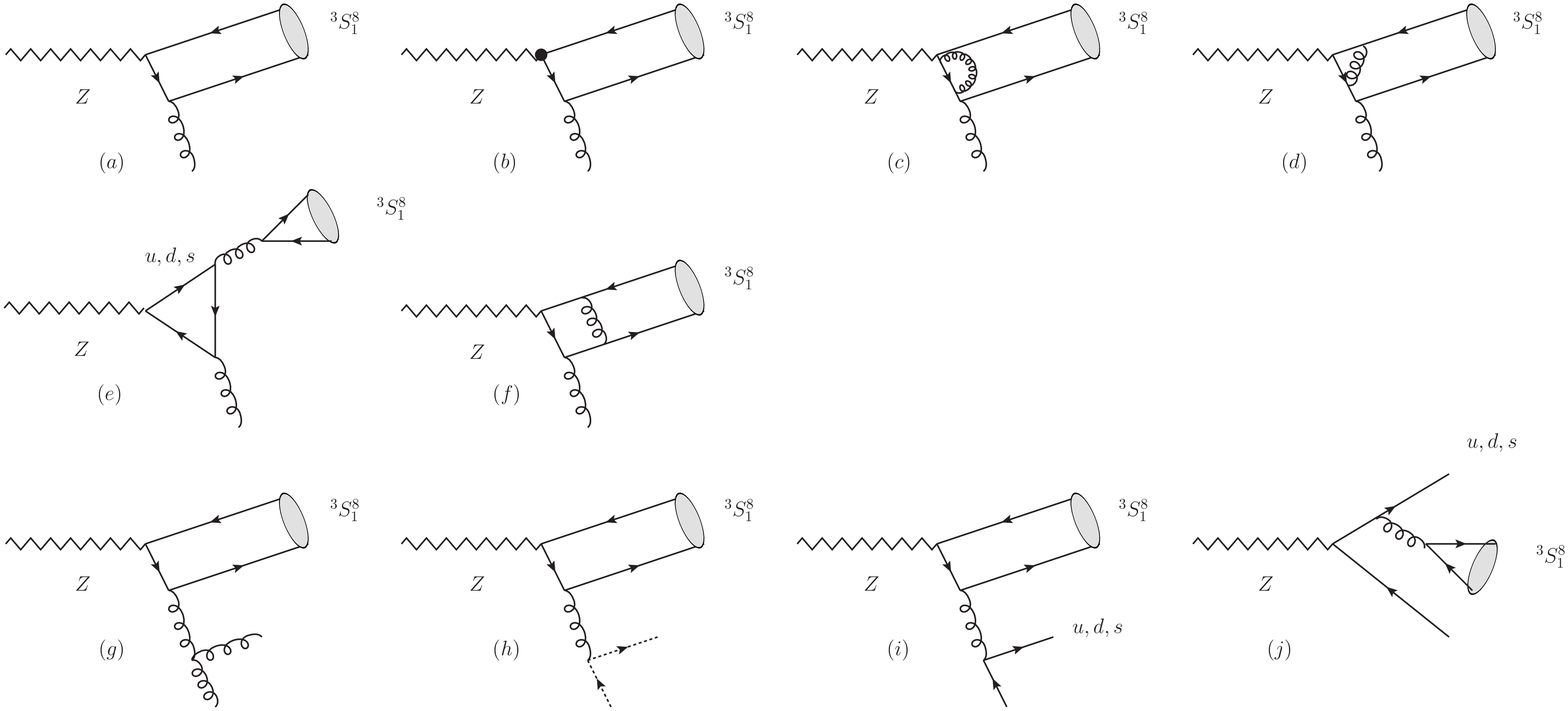}
\caption{\label{fig:Feyn1}
Some simple Feynman diagrams for the NLO processes of $^3S_1^{[8]}$.}
\end{figure*}

\begin{figure}
\includegraphics[width=0.6\textwidth]{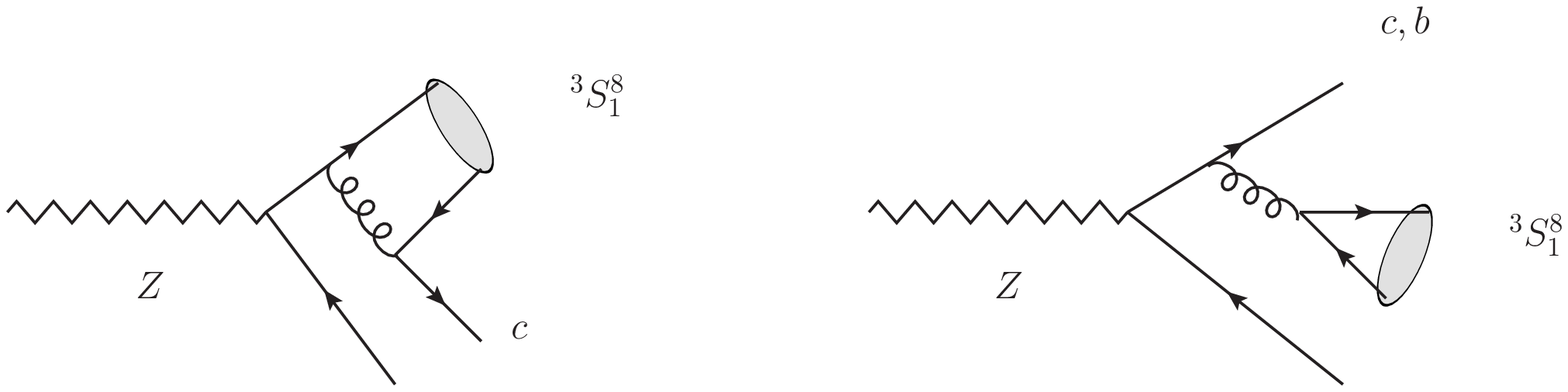}
\caption{\label{fig:Feyn2}
Some simple Feynman diagrams for the $\textrm{NLO}^{*}$ processes of $^3S_1^{[8]}$.}
\end{figure}

\begin{figure}
\includegraphics[width=0.6\textwidth]{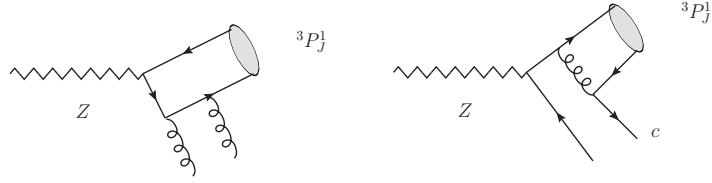}
\caption{\label{fig:Feyn3}
Some simple Feynman diagrams for the processes of $^3P_J^{[1]}$, including $Z \to c\bar{c}[^3P_J^{[1]}]+g+g$ and $Z \to c\bar{c}[^3P_J^{[1]}]+c+\bar{c}$.}
\end{figure}
Within the NRQCD framework, the decay width of $Z \to \chi_c(\chi_b)+X$ can be written as:
\begin{eqnarray}
d\Gamma=\sum_{n}d\hat{\Gamma}_{n}\langle \mathcal O ^{H}(n)\rangle,
\end{eqnarray}
where $d\hat{\Gamma}_n$ is the perturbative calculable short distance coefficients, representing the production of a configuration of the $Q\bar{Q}$ intermediate state with a quantum number $n(^{2S+1}L_J^{[1,8]})$. $\langle \mathcal O ^{H}(n)\rangle$ is the universal nonperturbative LDME. According to NRQCD, for $\chi_c$ and $\chi_b$ related processes, only two states should be taken into considerations at LO accuracy in $v$ , namely $^3S_1^{[8]}$ and $^3P_J^{[1]}$. Taking $\chi_c$ as an example, up to $\alpha\alpha_s^2$ order, for $n=^3S_1^{[8]}$ we have
\begin{eqnarray}
\textrm{LO}:&Z& \to c\bar{c}[^3S_1^{[8]}]+g, \nonumber \\
\textrm{NLO}:&Z& \to c\bar{c}[^3S_1^{[8]}]+g~(\textrm{virtual}), \nonumber \\
&Z& \to c\bar{c}[^3S_1^{[8]}]+g+g, \nonumber \\
&Z& \to c\bar{c}[^3S_1^{[8]}]+u_g+\bar{u}_g~(\textrm{ghost}), \nonumber \\
&Z& \to c\bar{c}[^3S_1^{[8]}]+u+\bar{u}, \nonumber \\
&Z& \to c\bar{c}[^3S_1^{[8]}]+d(s)+\bar{d}(\bar{s}), \nonumber \\
\textrm{NLO}^{*}:&Z& \to c\bar{c}[^3S_1^{[8]}]+c+\bar{c}, \nonumber \\
&Z& \to c\bar{c}[^3S_1^{[8]}]+b+\bar{b}. \label{3s18 channels}
\end{eqnarray}
The label ``$\textrm{NLO}^{*}$" represents the heavy quark-antiquark pair associated processes, which are free of divergence. In the case of $n=^3P_J^{[1]}$, there are two involved channels as listed below:
\begin{eqnarray}
&Z& \to c\bar{c}[^3P_J^{[1]}]+g+g, \nonumber \\
&Z& \to c\bar{c}[^3P_J^{[1]}]+c+\bar{c}. \label{3pj1 channels}
\end{eqnarray}
Some simple Feynman diagrams corresponding to Eqs. (\ref{3s18 channels}) and (\ref{3pj1 channels}) are presented in Figs. \ref{fig:Feyn1}, \ref{fig:Feyn2}, and \ref{fig:Feyn3}, including 51 diagrams for $^3S_1^{[8]}$ (2 LO diagrams, 6 counterterms, 15 one-loop, 18 diagrams for real corrections, and 10 NLO* diagrams), and 10 diagrams for $^3P_J^{[1]}$. Note that, as shown in Eq. (\ref{3s18 channels}), the real correction process $Z \to c\bar{c}[^3S_1^{[8]}]+q+\bar{q}$ has been divided into two categories, namely $q=u$ and $q=d(s)$. In addition, in Fig. 1(e) the diagrams involving fermion loops of $u,c$ and $d,s,b$ are also divided into two groups.

For the $\chi_b$ cases, one should replace the charm quark of Eqs. (\ref{3s18 channels}) and (\ref{3pj1 channels}) with the bottom quark. Of special attention is that the coupling of $Z c\bar{c}$ is different from $Zb\bar{b}$.

In the following, we will present the calculation formalisms for $Z \to Q\bar{Q}[^3S_1^{[8]}]+X$ and $Z \to Q\bar{Q}[^3P_J^{[1]}]+X$, respectively.

\subsection{$Z \to Q\bar{Q}[^3S_1^{[8]}]+X$}

To the next-to-leading order in $\alpha_s$, the decay width of $Z \to Q\bar{Q}[^3S_1^{[8]}]+X$ is
\begin{eqnarray}
\Gamma=\Gamma_{\textrm{Born}}+\Gamma_{\textrm{Virtual}}+\Gamma_{\textrm{Real}}+\mathcal O(\alpha\alpha_s^3),
\end{eqnarray}
where
\begin{eqnarray}
&&\Gamma_{\textrm{Virtual}}=\Gamma_{\textrm{Loop}}+\Gamma_{\textrm{CT}}, \nonumber \\
&&\Gamma_{\textrm{Real}}=\Gamma_{\textrm{S}}+\Gamma_{\textrm{HC}}+\Gamma_{\textrm{H}\overline{\textrm{C}}}.
\end{eqnarray}
$\Gamma_{\textrm{Virtual}}$ is the virtual corrections, consisting of the contributions from the one-loop diagrams ($\Gamma_{\textrm{Loop}}$) and the counterterms ($\Gamma_{\textrm{CT}}$). $\Gamma_{\textrm{Real}}$ means the real corrections, including the soft terms ($\Gamma_{S}$), hard-collinear terms $(\Gamma_{\textrm{HC}})$, and hard-noncollinear terms $(\Gamma_{\textrm{H}\overline{\textrm{C}}})$. For the purpose of isolating the ultraviolet (UV) and infrared (IR) divergences, we adopt the dimensional regularization with $D=4-2\epsilon$. The on-mass-shell (OS) scheme is employed to set the renormalization constants for the heavy quark mass ($Z_m$), heavy quark filed ($Z_2$), and gluon filed ($Z_3$). The modified minimal-subtraction ($\overline{MS}$) scheme is for the QCD gauge coupling ($Z_g$), as listed below ($Q=c,b$) \cite{Klasen:2004tz}
\begin{eqnarray}
\delta Z_{m}^{OS}&=& -3 C_{F} \frac{\alpha_s N_{\epsilon}}{4\pi}\left[\frac{1}{\epsilon_{\textrm{UV}}}-\gamma_{E}+\textrm{ln}\frac{4 \pi \mu_r^2}{m_Q^2}+\frac{4}{3}+\mathcal O(\epsilon)\right], \nonumber \\
\delta Z_{2}^{OS}&=& - C_{F} \frac{\alpha_s N_{\epsilon}}{4\pi}\left[\frac{1}{\epsilon_{\textrm{UV}}}+\frac{2}{\epsilon_{\textrm{IR}}}-3 \gamma_{E}+3 \textrm{ln}\frac{4 \pi \mu_r^2}{m_Q^2} \right. \nonumber\\
&& \left.+4+\mathcal O(\epsilon)\right], \nonumber \\
\delta Z_{3}^{\overline{MS}}&=& \frac{\alpha_s N_{\epsilon}}{4\pi}\left[\beta_{0}(n_{lf})-2 C_{A}\right]\left[(\frac{1}{\epsilon_{\textrm{UV}}}-\frac{1}{\epsilon_{\textrm{IR}}})  \right. \nonumber\\
&& \left. -\frac{4}{3}T_F(\frac{1}{\epsilon_{\textrm{UV}}}-\gamma_E+\textrm{ln}\frac{4\pi\mu_r^2}{m_c^2}) \right. \nonumber\\
&& \left. -\frac{4}{3}T_F(\frac{1}{\epsilon_{\textrm{UV}}}-\gamma_E+\textrm{ln}\frac{4\pi\mu_r^2}{m_b^2})+\mathcal O(\epsilon)\right], \nonumber \\
\delta Z_{g}^{\overline{MS}}&=& -\frac{\beta_{0}(n_f)}{2}\frac{\alpha_s N_{\epsilon}}{4\pi}\left[\frac{1} {\epsilon_{\textrm{UV}}}-\gamma_{E}+\textrm{ln}(4\pi)+\mathcal O(\epsilon)\right], \label{CT}
\end{eqnarray}
where $\gamma_E$ is the Euler's constant, $\beta_{0}(n_f)=\frac{11}{3}C_A-\frac{4}{3}T_Fn_f$ is the one-loop coefficient of the $\beta$-function, and $\beta_{0}(n_{lf})$ is identical to $\frac{11}{3}C_A-\frac{4}{3}T_Fn_{lf}$. $n_f$ and $n_{lf}$ are the number of active quark flavors and light quark flavors, respectively. $N_{\epsilon}= \Gamma[1-\epsilon] /({4\pi\mu_r^2}/{(4m_c^2)})^{\epsilon}$. In ${\rm SU}(3)_c$, the color factors are given by $T_F=\frac{1}{2}$, $C_F=\frac{4}{3}$, and $C_A=3$. To subtract the IR divergences in $\Gamma_{\textrm{Real}}$, the two-cutoff slicing strategy \cite{Harris:2001sx} is utilized.

To calculate the D-dimension trace of the fermion loop involving $\gamma_5$, under the scheme described in \cite{Korner:1991sx}, we write down all the amplitudes from the same starting point (such as the $Z$-vertex) and abandon the cyclicity. As a crosscheck for the correctness of the treatments on $\gamma_5$, we have calculated the QCD NLO corrections to the similar process, $Z \to c\bar{c}[^3S_1^{[1]}]+\gamma$, obtaining exactly the same $K$ factor as in \cite{Wang:2013ywc}.

\subsection{$Z \to Q\bar{Q}[^3P_J^{[1]}]+X$}

The heavy quark-antiquark associated process $Z \to Q\bar{Q}[^3P_J^{[1]}]+Q+\bar{Q}$ ($Q=c,b$) is finite, thus one can calculate it directly. Now we are to deal with the other process of $Z \to Q\bar{Q}[^3P_J^{[1]}]+g+g$ ($Q=c,b$), which has soft singularities. Taking $\chi_c$ as an example, we first divide $\Gamma(Z \to c\bar{c}[^3P_J^{[1]}]+g+g)$ into two terms,
\begin{eqnarray}
&&d\Gamma(Z \to c\bar{c}[^3P_J^{[1]}]+g+g)=d\hat{\Gamma}_{^3P_J^{[1]}} \langle \mathcal O^{\chi_c}(^3P_J^{[1]}) \rangle+d\hat{\Gamma}_{^3S_1^{[8]}}^{LO} \langle \mathcal O^{\chi_c}(^3S_1^{[8]}) \rangle ^{NLO}.
\end{eqnarray}
Then we have
\begin{eqnarray}
d\hat{\Gamma}_{^3P_J^{[1]}} \langle \mathcal O^{\chi_c}(^3P_J^{[1]}) \rangle&=&d\Gamma(Z \to c\bar{c}[^3P_J^{[1]}]+g+g) -d\hat{\Gamma}_{^3S_1^{[8]}}^{LO} \langle \mathcal O^{\chi_c}(^3S_1^{[8]}) \rangle ^{NLO} \nonumber \\
&=&d{\Gamma}_F+(d{\Gamma}_S-d\hat{\Gamma}_{^3S_1^{[8]}}^{LO} \langle \mathcal O^{\chi_c}(^3S_1^{[8]}) \rangle ^{NLO}) \nonumber \\
&=&d{\Gamma}_F+d{\Gamma}^{*}. \label{3pj1 SDC}
\end{eqnarray}
$d{\Gamma}^{*}$ denotes the sum of $d{\Gamma}_S$ and $-d\hat{\Gamma}_{^3S_1^{[8]}}^{LO} \langle \mathcal O^{\chi_c}(^3S_1^{[8]}) \rangle ^{NLO}$. $d\Gamma_F$ is the finite terms in $d\Gamma(Z \to c\bar{c}[^3P_J^{[1]}]+g+g)$, and $d\Gamma_S$ is the soft part which can be written as
\begin{eqnarray}
&&d{\Gamma}_S=-\frac{\alpha_s}{3 \pi m_c} u^{s}_\epsilon \frac{N_c^2-1}{N_c} d\hat{\Gamma}^{LO}_{^3S_1^{[8]}} \langle \mathcal O^{\chi_c}(^3P_J^{[1]}) \rangle, \label{3pj1 soft}
\end{eqnarray}
with
\begin{eqnarray}
u^{s}_\epsilon=\frac{1}{\epsilon_{IR}}+\frac{E}{|\textbf{p}|} \textrm{ln}(\frac{E+|\textbf{p}|}{E-|\textbf{p}|}) + \textrm{ln}(\frac{4 \pi \mu_r^2}{s\delta_s^2})-\gamma_E-\frac{1}{3}. \label{us}
\end{eqnarray}
$N_c$ is identical to 3 for $SU(3)$ gauge field. $E$ and $\textbf{p}$ denote the energy and 3-momentum of $\chi_c$, respectively. $\delta_s$ is the usual ``soft cut" employed to impose an amputation on the energy of the emitted gluon.

Now we are to calculate the transition rate of $^3S_1^{[8]}$ into $^3P_J^{[1]}$. From Ref. \cite{Jia:2014jfa}, under the dimensional regularization scheme we have
\begin{eqnarray}
\langle \mathcal O^{\chi_c}(^3S_1^{[8]}) \rangle ^{NLO}=-\frac{\alpha_s}{3 \pi m_c} u^{c}_\epsilon \frac{N_c^2-1}{N_c} \langle \mathcal O^{\chi_c}(^3P_J^{[1]}) \rangle. \label{3s18to3pj1}\
\end{eqnarray}
On the basis of $\mu_{\Lambda}$-cutoff scheme \cite{Jia:2014jfa}, $u^{c}_\epsilon$ has the form of
\begin{eqnarray}
u^{s}_\epsilon=\frac{1}{\epsilon_{IR}}-\gamma_E-\frac{1}{3}- \textrm{ln}(\frac{4 \pi \mu_r^2}{\mu_{\Lambda}^2}). \label{uc}
\end{eqnarray}
$\mu_{\Lambda}$ is the upper bound of the integrated gluon energy, rising from the renormalization of the LDME. Substituting Eqs. (\ref{3pj1 soft}), (\ref{us}), (\ref{3s18to3pj1}), and (\ref{uc}) into Eq. (\ref{3pj1 SDC}), the soft singularities in $d{\Gamma}_S$ and $d\hat{\Gamma}_{^3S_1^{[8]}}^{LO} \langle \mathcal O^{\chi_c}(^3S_1^{[8]}) \rangle ^{NLO}$ cancel each other. Consequently $d{\Gamma}^{*}$ is free of divergence.

For the $\chi_b$ cases, one should replace the charm quark with the bottom quark. In addition, the $Zc\bar{c}$ coupling should changed into the $Zb\bar{b}$ form.

\section{Numerical results and discussions}

Before presenting the phenomenological results, we first demonstrate the choices of the parameters in our calculations. To keep the gauge invariance, the masses of $\chi_c$ and $\chi_b$ are set to be $2m_c$ and $2m_b$, respectively. $m_c=1.5 \pm 0.1$ GeV and $m_b=4.9 \pm 0.2$ GeV. $m_Z=91.1876$ GeV. $\alpha=1/137$. In the calculations for the NLO, the $\textrm{NLO}^{*}$, and the two $^3P_J^{[1]}$ processes, we employ the two-loop $\alpha_s$ running, and one-loop $\alpha_s$ running for LO. We take $m_c(m_b)$ as the value of $\mu_{\Lambda}$ for $\chi_c(\chi_{b})$. The values of $\langle \mathcal O^{\chi_{c}(\chi_{b})}(^3S_1^8) \rangle$ are taken as
\begin{eqnarray}
&&\langle \mathcal O^{\chi_{c0}}(^3S_1^{[8]}) \rangle=2.15 \times 10^{-3}~\textrm{GeV}^3, \nonumber \\
&&\langle \mathcal O^{\chi_{b0}}(^3S_1^{[8]}) \rangle=9.40 \times 10^{-3}~\textrm{GeV}^3,
\end{eqnarray}
from Refs. \cite{Feng:2015wka} and \cite{Jia:2014jfa}. In the case of the $^3P_J^{[1]}$ channels, the relation $\langle \mathcal O^{\chi_{cJ}(\chi_{bJ})}(^3P_J^{[1]}) \rangle=\frac{9}{2\pi}(2J+1)|R^{'}_p(0)|^2$ is adopted with $|R^{'}_p(0)|^2=0.075~\textrm{GeV}^5$ for $\chi_c$ and $|R^{'}_p(0)|^2=1.417~\textrm{GeV}^5$ for $\chi_b$.

\begin{figure*}
\includegraphics[width=0.49\textwidth]{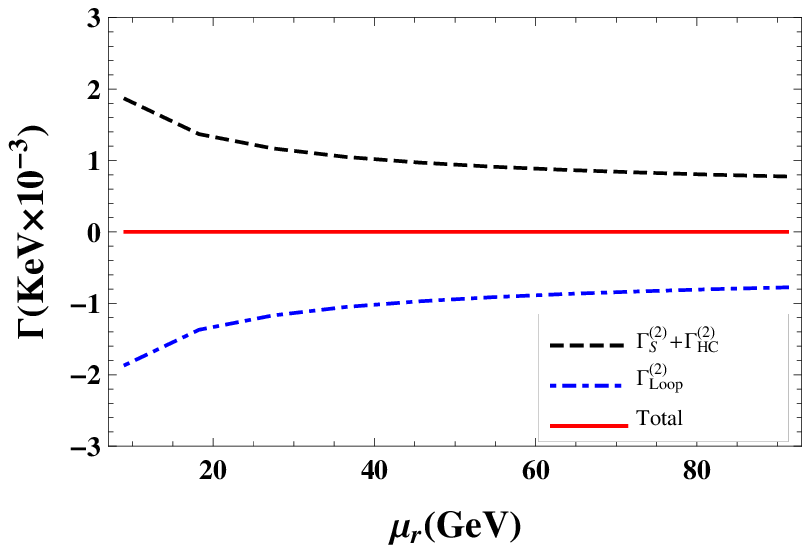}
\includegraphics[width=0.49\textwidth]{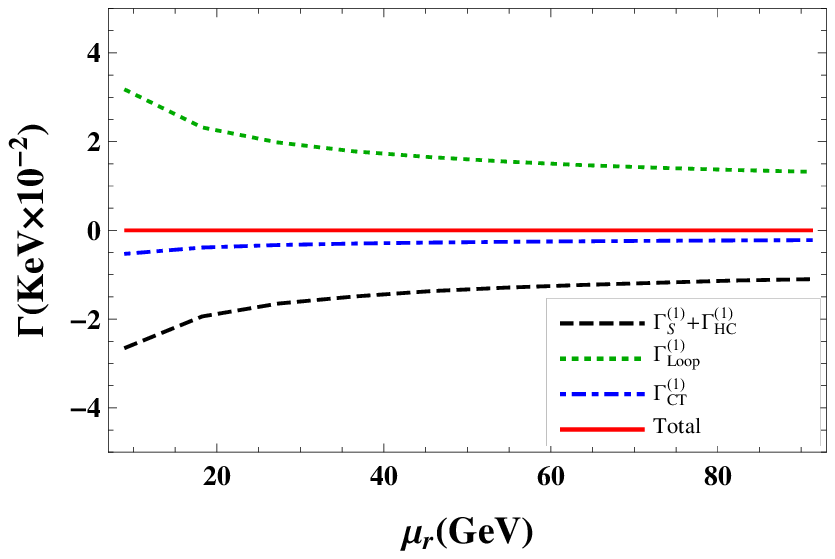}
\caption{\label{fig:div}
Cancellation of the $\epsilon^{-2}$- and $\epsilon^{-1}$- order divergences for $Z \to c\bar{c}[^3S_1^{[8]}]+X$. The superscripts ``(2)" and ``(1)" denote the $\epsilon^{-2}$- and $\epsilon^{-1}$- order terms, respectively.}
\end{figure*}

\begin{figure*}
\includegraphics[width=0.49\textwidth]{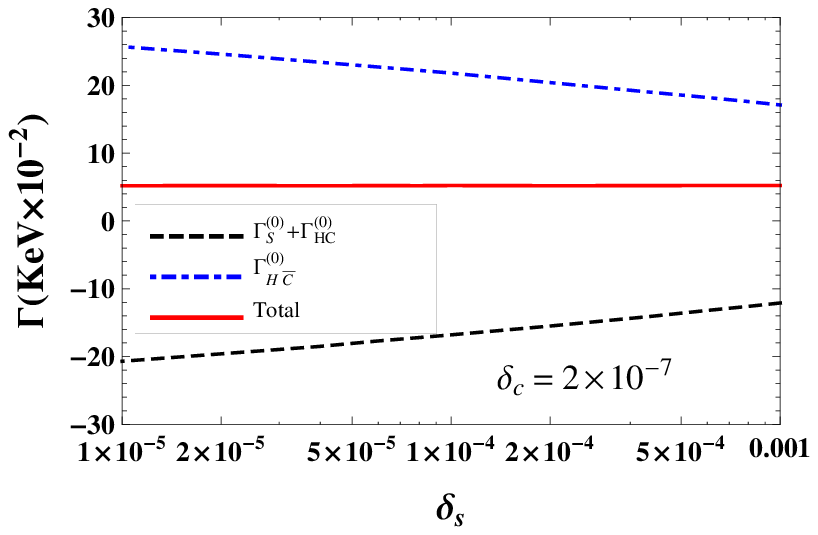}
\includegraphics[width=0.49\textwidth]{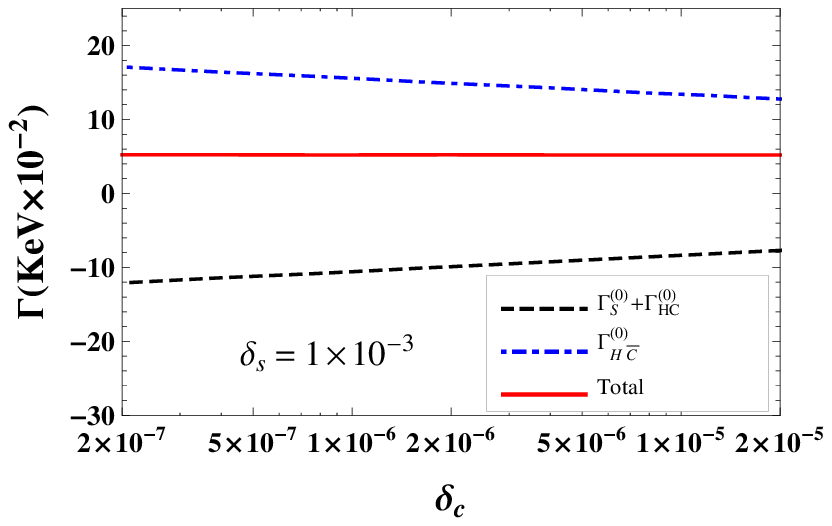}
\caption{\label{fig:cut3s18}
The verification of the independence on the cutoff parameters of $\delta_s$ and $\delta_c$ for $Z \to c\bar{c}[^3S_1^{[8]}]+X$. The superscript ``(0)" denotes the $\epsilon^{0}$-order terms.}
\end{figure*}

\begin{figure*}
\includegraphics[width=0.49\textwidth]{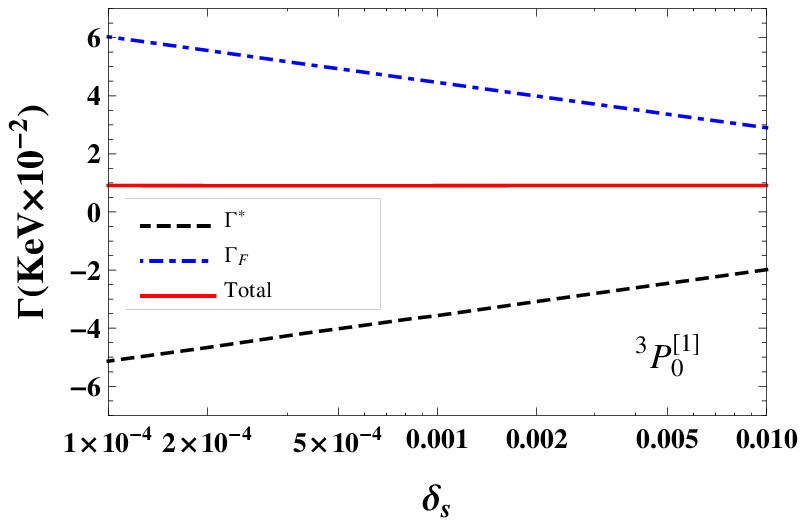}
\includegraphics[width=0.49\textwidth]{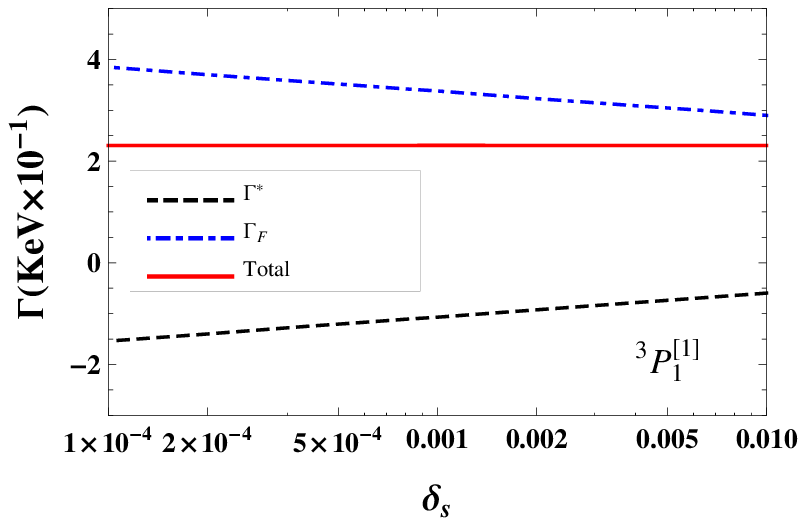}
\includegraphics[width=0.49\textwidth]{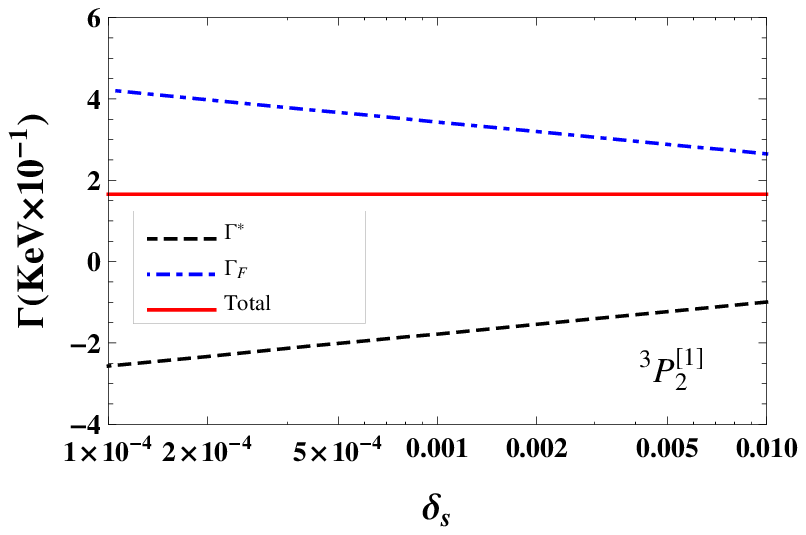}
\caption{\label{fig:cut3pj1}
The verification of the independence on the cutoff parameter of $\delta_s$ for $Z \to c\bar{c}[^3P_J^{[1]}]+g+g$.}
\end{figure*}

In our calculations, the mathematica package $\textbf{Malt@FDC}$~\cite{Feng:2017bdu, Sun:2017nly, Sun:2017wxk,Sun:2018rgx} is employed to obtain $\Gamma_{\textrm{Virtual}}$, $\Gamma_{\textrm{\textrm{S}}}$ and $\Gamma_{\textrm{\textrm{HC}}}$. $\textbf{FDC}$~\cite{Wang:2004du} package serves as the agent to evaluate the contributions of the hard-noncollinear part of the real corrections, namely $\Gamma_{\textrm{H}\overline{\textrm{C}}}$.  Both the cancellation of divergence and the independence on cutoff have been checked carefully. By taking $\chi_c$ as an example, we present the verifications in Figs. \ref{fig:div}, \ref{fig:cut3s18}, and \ref{fig:cut3pj1}. Note that, for $Z \to c\bar{c}[^3S_1^{[8]}]+q+\bar{q}$ (as displayed in Figures. \ref{fig:Feyn1}(i) and \ref{fig:Feyn1}(j)), the contributions of the single-gluon-fragmentation (SGF) diagrams (\ref{fig:Feyn1}(j)) are free of divergence. Moreover, the SGF contribution is about 2 orders of magnitude bigger than that of Fig. \ref{fig:Feyn1}(i). In order to clearly demonstrate the verification of the independence on the cutoff parameters ($\delta_s,\delta_c$), the $\Gamma_{\textrm{H}\overline{\textrm{C}}}$ in Fig. \ref{fig:cut3s18} does not include the SGF contributions.

\subsection{Phenomenological results for $\chi_c$}

The NRQCD predictions for $\Gamma(Z \to \chi_{cJ}+X)$ ($J=0,1,2$) are demonstrated in Tables. \ref{xc0}, \ref{xc1}, and \ref{xc2}, respectively.
\begin{table*}[htb]
\caption{The decay widths (unit: KeV) of $\Gamma(Z \to \chi_{c0}+X)$. $\mu_{\Lambda}=m_c$.}
\label{xc0}
\begin{tabular}{ccccccccc}
\hline
$\mu_r$ & $m_c(\textrm{GeV})$ & $^3S_1^{[8]}|_{\textrm{LO}}$ & $^3S_1^{[8]}|_{\textrm{NLO}}$ & $^3S_1^{[8]}|_{\textrm{NLO}^{*}}$ & $^3P_0^{[1]}|_{gg}$ & $^3P_0^{[1]}|_{c\bar{c}}$ & $\Gamma_{\textrm{total}}$ & $\textrm{Br}(10^{-5})$\\ \hline
$~$ & $1.4$ & $1.20 \times 10^{-2}$ & $14.9$ & $8.26$ & $5.63 \times 10^{-2}$ & $27.0$ & $50.2$ & $2.02$\\
$2m_c$ & $1.5$ & $1.09 \times 10^{-2}$ & $10.9$ & $6.05$ & $4.27 \times 10^{-2}$ & $18.1$ & $35.1$ & $1.41$\\
$~$ & $1.6$ & $9.99 \times 10^{-3}$ & $8.12$ & $4.53$ & $3.30 \times 10^{-2}$ & $12.5$ & $25.1$ & $1.01$\\ \hline
$~$ & $1.4$ & $5.30 \times 10^{-3}$ & $2.99$ & $1.66$ & $1.13 \times 10^{-2}$ & $5.43$ & $10.1$ & $0.41$\\
$m_Z$ & $1.5$ & $4.95 \times 10^{-3}$ & $2.31$ & $1.28$ & $9.06 \times 10^{-3}$ & $3.84$ & $7.45$ & $0.30$\\
$~$ & $1.6$ & $4.64 \times 10^{-3}$ & $1.82$ & $1.01$ & $7.36 \times 10^{-3}$ & $2.78$ & $5.61$ & $0.23$\\ \hline
\end{tabular}
\end{table*}
\begin{table*}[htb]
\caption{The decay widths (unit: KeV) of $\Gamma(Z \to \chi_{c1}+X)$. $\mu_{\Lambda}=m_c$.}
\label{xc1}
\begin{tabular}{ccccccccc}
\hline
$\mu_r$ & $m_c(\textrm{GeV})$ & $^3S_1^{[8]}|_{\textrm{LO}}$ & $^3S_1^{[8]}|_{\textrm{NLO}}$ & $^3S_1^{[8]}|_{\textrm{NLO}^{*}}$ & $^3P_1^{[1]}|_{gg}$ & $^3P_1^{[1]}|_{c\bar{c}}$ & $\Gamma_{\textrm{total}}$ & $\textrm{Br}(10^{-5})$\\ \hline
$~$ & $1.4$ & $3.60 \times 10^{-2}$ & $44.6$ & $24.8$ & $1.47$ & $29.9$ & $101$ & $4.06$\\
$2m_c$ & $1.5$ & $3.27 \times 10^{-2}$ & $32.6$ & $18.2$ & $1.09$ & $20.0$ & $71.9$ & $2.89$\\
$~$ & $1.6$ & $3.00 \times 10^{-2}$ & $24.4$ & $13.6$ & $0.819$ & $13.7$ & $52.5$ & $2.11$\\ \hline
$~$ & $1.4$ & $1.59 \times 10^{-2}$ & $8.98$ & $4.98$ & $0.296$ & $6.01$ & $20.3$ & $0.82$\\
$m_Z$ & $1.5$ & $1.49 \times 10^{-2}$ & $6.94$ & $3.85$ & $0.231$ & $4.23$ & $15.3$ & $0.61$\\
$~$ & $1.6$ & $1.39 \times 10^{-2}$ & $5.45$ & $3.03$ & $0.183$ & $3.05$ & $11.7$ & $0.47$\\ \hline
\end{tabular}
\end{table*}
\begin{table*}[htb]
\caption{The decay widths (unit: KeV) of $\Gamma(Z \to \chi_{c2}+X)$. $\mu_{\Lambda}=m_c$.}
\label{xc2}
\begin{tabular}{ccccccccc}
\hline
$\mu_r$ & $m_c(\textrm{GeV})$ & $^3S_1^{[8]}|_{\textrm{LO}}$ & $^3S_1^{[8]}|_{\textrm{NLO}}$ & $^3S_1^{[8]}|_{\textrm{NLO}^{*}}$ & $^3P_2^{[1]}|_{gg}$ & $^3P_2^{[1]}|_{c\bar{c}}$ & $\Gamma_{\textrm{total}}$ & $\textrm{Br}(10^{-5})$\\ \hline
$~$ & $1.4$ & $6.00 \times 10^{-2}$ & $74.3$ & $41.3$ & $1.03$ & $11.7$ & $128$ & $5.14$\\
$2m_c$ & $1.5$ & $5.46 \times 10^{-2}$ & $54.4$ & $30.3$ & $0.780$ & $7.84$ & $93.2$ & $3.74$\\
$~$ & $1.6$ & $4.99 \times 10^{-2}$ & $40.6$ & $22.6$ & $0.601$ & $5.39$ & $69.2$ & $2.78$\\ \hline
$~$ & $1.4$ & $2.65 \times 10^{-2}$ & $15.0$ & $8.30$ & $0.208$ & $2.35$ & $25.8$ & $1.04$\\
$m_Z$ & $1.5$ & $2.48 \times 10^{-2}$ & $11.6$ & $6.42$ & $0.166$ & $1.66$ & $19.8$ & $0.80$\\
$~$ & $1.6$ & $2.32 \times 10^{-2}$ & $9.08$ & $5.05$ & $0.134$ & $1.20$ & $15.5$ & $0.62$\\ \hline
\end{tabular}
\end{table*}
One can see that the branching rations are on the order of $10^{-5}$, indicating a detectable prospect of these decay processes at LHC or other platforms. To be specific, considering the uncertainties induced by the choices of the values of $\mu_r(2m_c \sim M_Z)$ and $m_c(1.4\ \sim 1.6~\textrm{GeV})$, we have
\begin{eqnarray}
\textrm{Br}(Z \to \chi_{c0}+X)&=&(0.23 - 2.02) \times 10^{-5}, \nonumber \\
\textrm{Br}(Z \to \chi_{c1}+X)&=&(0.47 - 4.06) \times 10^{-5}, \nonumber \\
\textrm{Br}(Z \to \chi_{c2}+X)&=&(0.62 - 5.14) \times 10^{-5}.
\end{eqnarray}

For the color-singlet $^3P_J^{[1]}$ ($J=0,1,2$) state cases, the process of $Z \to c\bar{c}[^3P_J^{[1]}]+c+\bar{c}$ serves as the leading role in the total CS prediction, due to the $c$-quark fragmentation mechanism. The other CS process, namely $Z \to c\bar{c}[^3P_J^{[1]}]+g+g$, contributes moderately, accounting for about $0.24\%,5\%$, and $10\%$ of the total CS prediction for $J=0,1,2$, respectively.

In the case of the color-octet $^3S_1^{[8]}$ state, the QCD NLO corrections can enhance the LO results significantly, by $2 - 3$ orders. This can be attributed to the kinematic enhancements via the $^3S_1^{[8]}$ single-gluon-fragmentation diagrams, including the one-loop triangle anomalous diagrams (Fig. \ref{fig:Feyn1}(e)) and the diagrams associated with a final $q\bar{q}$ ($q=u,d,s$) pair (Fig. \ref{fig:Feyn1}(j)), which first emerge at the NLO level. By the same token, the $\textrm{NLO}^{*}$ channels can also provide considerable contributions, about one half of the NLO results. Consequently the CO channels will play a vital role in the decay process of $Z \to \chi_c+X$. To show the CO significance obviously, we introduce the following ratios
\begin{eqnarray}
\Gamma^{\chi_{c0}}_{\textrm{CO}} / \Gamma^{\chi_{c0}}_{\textrm{\textrm{CS+CO}}}&=&(46.1 - 50.3)\%, \nonumber \\
\Gamma^{\chi_{c1}}_{\textrm{CO}} / \Gamma^{\chi_{c1}}_{\textrm{\textrm{CS+CO}}}&=&(68.9 - 72.4)\%, \nonumber \\
\Gamma^{\chi_{c2}}_{\textrm{CO}} / \Gamma^{\chi_{c2}}_{\textrm{\textrm{CS+CO}}}&=&(90.1 - 91.4)\%.
\end{eqnarray}
$\Gamma^{\chi_{cJ}}_{\textrm{CO}}$ ($J=0,1,2$) denotes the sum of the NLO and $\textrm{NLO}^{*}$ result.

In addition to the crucial impacts on the total widths, the CO channels can also significantly influence the predictions on the ratios of $\Gamma_{\chi_{c1}}/\Gamma_{\chi_{c0}}$ and $\Gamma_{\chi_{c2}}/\Gamma_{\chi_{c0}}$, as shown below
\begin{eqnarray}
\textrm{CS}&:&~~~\Gamma_{\chi_{c1}} / \Gamma_{\chi_{c0}} = 1.159 - 1.162, \nonumber \\
\textrm{CS+CO}&:&~~~\Gamma_{\chi_{c1}} / \Gamma_{\chi_{c0}} = 2.007 - 2.087, \nonumber \\
\textrm{CS}&:&~~~\Gamma_{\chi_{c2}} / \Gamma_{\chi_{c0}} = 0.471 - 0.480, \nonumber \\
\textrm{CS+CO}&:&~~~\Gamma_{\chi_{c2}} / \Gamma_{\chi_{c0}} = 2.558 - 2.756.
\end{eqnarray}
One can see that the CS results have been thoroughly changed by including the CO states. The conspicuous differences can be regarded as an outstanding probe to distinguish between the CO and CS mechanism.

Considering the branching ratios of $\chi_c $ to $ J/\psi$ are not small \cite{Tanabashi:2018oca},
\begin{eqnarray}
\textrm{Br}(\chi_{c0} \to J/\psi+\gamma)&=&1.4\%, \nonumber \\
\textrm{Br}(\chi_{c1} \to J/\psi+\gamma)&=&34.3\%, \nonumber \\
\textrm{Br}(\chi_{c2} \to J/\psi+\gamma)&=&19.0\%,
\end{eqnarray}
thus $\chi_c$ feeddown may have a substantial impact on the production of $J/\psi$. Adding together the contributions from $\chi_{c0}$, $\chi_{c1}$, and $\chi_{c2}$, we finally obtain
\begin{eqnarray}
\Gamma(Z \to J/\psi+X)|_{\chi_c-\textrm{feeddown}}=(0.28 \sim 2.4) \times 10^{-5}.
\end{eqnarray}
This result is about one order of magnitude smaller than the experimental data released from the L3 Collaboration at LEP \cite{Acciarri:1998iy}.

\subsection{Phenomenological results for $\chi_b$}
\begin{table*}[htb]
\caption{The decay widths (unit: KeV) of $\Gamma(Z \to \chi_{b0}+X)$. $\mu_{\Lambda}=m_b$.}
\label{xb0}
\begin{tabular}{ccccccccc}
\hline
$\mu_r$ & $m_b(\textrm{GeV})$ & $^3S_1^{[8]}|_{\textrm{LO}}$ & $^3S_1^{[8]}|_{\textrm{NLO}}$ & $^3S_1^{[8]}|_{\textrm{NLO}^{*}}$ & $^3P_0^{[1]}|_{gg}$ & $^3P_0^{[1]}|_{b\bar{b}}$ & $\Gamma_{\textrm{total}}$ & $\textrm{Br}(10^{-7})$\\ \hline
$~$ & $4.7$ & $9.76 \times 10^{-3}$ & $0.272$ & $0.148$ & $8.84 \times 10^{-3}$ & $0.677$ & $1.11$ & $4.46$\\
$2m_b$ & $4.9$ & $9.26 \times 10^{-3}$ & $0.225$ & $0.121$ & $7.46 \times 10^{-3}$ & $0.535$ & $0.888$ & $3.57$\\
$~$ & $5.1$ & $8.82 \times 10^{-3}$ & $0.187$ & $9.95 \times 10^{-2}$ & $6.34 \times 10^{-3}$ & $0.426$ & $0.719$ & $2.89$\\ \hline
$~$ & $4.7$ & $6.34 \times 10^{-3}$ & $0.119$ & $6.26 \times 10^{-2}$ & $3.74 \times 10^{-3}$ & $0.286$ & $0.472$ & $1.90$\\
$m_Z$ & $4.9$ & $6.08 \times 10^{-3}$ & $0.101$ & $5.22 \times 10^{-2}$ & $3.22 \times 10^{-3}$ & $0.231$ & $0.387$ & $1.55$\\
$~$ & $5.1$ & $5.85 \times 10^{-3}$ & $8.62 \times 10^{-2}$ & $4.37 \times 10^{-2}$ & $2.79 \times 10^{-3}$ & $0.187$ & $0.320$ & $1.29$\\ \hline
\end{tabular}
\end{table*}
\begin{table*}[htb]
\caption{The decay widths (unit: KeV) of $\Gamma(Z \to \chi_{b1}+X)$. $\mu_{\Lambda}=m_b$.}
\label{xb1}
\begin{tabular}{ccccccccc}
\hline
$\mu_r$ & $m_b(\textrm{GeV})$ & $^3S_1^{[8]}|_{\textrm{LO}}$ & $^3S_1^{[8]}|_{\textrm{NLO}}$ & $^3S_1^{[8]}|_{\textrm{NLO}^{*}}$ & $^3P_1^{[1]}|_{gg}$ & $^3P_1^{[1]}|_{b\bar{b}}$ & $\Gamma_{\textrm{total}}$ & $\textrm{Br}(10^{-7})$\\ \hline
$~$ & $4.7$ & $2.92 \times 10^{-2}$ & $0.814$ & $0.445$ & $0.153$ & $0.653$ & $2.06$ & $8.28$\\
$2m_b$ & $4.9$ & $2.78 \times 10^{-2}$ & $0.674$ & $0.362$ & $0.128$ & $0.512$ & $1.68$ & $6.74$\\
$~$ & $5.1$ & $2.64 \times 10^{-2}$ & $0.562$ & $0.299$ & $0.109$ & $0.405$ & $1.37$ & $5.50$\\ \hline
$~$ & $4.7$ & $1.90 \times 10^{-2}$ & $0.357$ & $0.188$ & $6.47 \times 10^{-2}$ & $0.276$ & $0.886$ & $3.56$\\
$m_Z$ & $4.9$ & $1.83 \times 10^{-2}$ & $0.303$ & $0.157$ & $5.54 \times 10^{-2}$ & $0.221$ & $0.736$ & $2.96$\\
$~$ & $5.1$ & $1.75 \times 10^{-2}$ & $0.258$ & $0.131$ & $4.77 \times 10^{-2}$ & $0.178$ & $0.616$ & $2.47$\\ \hline
\end{tabular}
\end{table*}
\begin{table*}[htb]
\caption{The decay widths (unit: KeV) of $\Gamma(Z \to \chi_{b2}+X)$. $\mu_{\Lambda}=m_b$.}
\label{xb2}
\begin{tabular}{ccccccccc}
\hline
$\mu_r$ & $m_b(\textrm{GeV})$ & $^3S_1^{[8]}|_{\textrm{LO}}$ & $^3S_1^{[8]}|_{\textrm{NLO}}$ & $^3S_1^{[8]}|_{\textrm{NLO}^{*}}$ & $^3P_2^{[1]}|_{gg}$ & $^3P_2^{[1]}|_{b\bar{b}}$ & $\Gamma_{\textrm{total}}$ & $\textrm{Br}(10^{-7})$\\ \hline
$~$ & $4.7$ & $4.88 \times 10^{-2}$ & $1.360$ & $0.743$ & $0.160$ & $0.270$ & $2.53$ & $10.2$\\
$2m_b$ & $4.9$ & $4.63 \times 10^{-2}$ & $1.130$ & $0.605$ & $0.136$ & $0.212$ & $2.08$ & $8.35$\\
$~$ & $5.1$ & $4.41 \times 10^{-2}$ & $0.937$ & $0.497$ & $0.116$ & $0.168$ & $1.72$ & $6.91$\\ \hline
$~$ & $4.7$ & $3.18 \times 10^{-2}$ & $0.595$ & $0.312$ & $6.78 \times 10^{-2}$ & $0.114$ & $1.09$ & $4.38$\\
$m_Z$ & $4.9$ & $3.04 \times 10^{-2}$ & $0.504$ & $0.261$ & $5.87 \times 10^{-2}$ & $9.12 \times 10^{-2}$ & $0.915$ & $3.69$\\
$~$ & $5.1$ & $2.92 \times 10^{-2}$ & $0.430$ & $0.219$ & $5.12 \times 10^{-2}$ & $7.36 \times 10^{-2}$ & $0.774$ & $3.11$\\ \hline
\end{tabular}
\end{table*}
Based on NRQCD, the predicted decay widths via $Z \to \chi_{bJ}+X$ ($J=0,1,2$) are presented in Tables. \ref{xb0}, \ref{xb1}, and \ref{xb2}. It is observed that the branching ratio for $Z \to \chi_{bJ}+X$ is around $10^{-7} - 10^{-6}$. Taking into account the uncertainties induced by $\mu_r$ ($2m_b \sim M_Z$) and the mass of $b$ quark ($4.7 \sim 5.1$ GeV), we have
\begin{eqnarray}
\textrm{Br}(Z \to \chi_{b0}+X)&=&(1.29 - 4.46) \times 10^{-7}, \nonumber \\
\textrm{Br}(Z \to \chi_{b1}+X)&=&(2.47 - 8.28) \times 10^{-7}, \nonumber \\
\textrm{Br}(Z \to \chi_{b2}+X)&=&(0.31 - 1.02) \times 10^{-6}.
\end{eqnarray}

In contrast to the previously stated ``moderation" of the contributions via $Z \to c\bar{c}[^3P_J^{[1]}]+g+g$, the channel $Z \to b\bar{b}[^3P_J^{[1]}]+g+g$ contributes significantly,
\begin{eqnarray}
&&\Gamma_{^3P_0^{[1]}}^{gg}/\Gamma_{^3P_0^{[1]}}^{\textrm{CS}} \sim 1.5\%, \nonumber \\
&&\Gamma_{^3P_1^{[1]}}^{gg}/\Gamma_{^3P_1^{[1]}}^{\textrm{CS}} \sim 20\%, \nonumber \\
&&\Gamma_{^3P_2^{[1]}}^{gg}/\Gamma_{^3P_2^{[1]}}^{\textrm{CS}} \sim 40\%.
\end{eqnarray}
$\Gamma_{^3P_J^{[1]}}^{gg}$ means $\Gamma(Z \to b\bar{b}[{^3P_J^{[1]}}]+g+g)$. $\Gamma_{^3P_J^{[1]}}^{\textrm{CS}}$ is the total color-singlet predictions, including both $\Gamma(Z \to b\bar{b}[{^3P_J^{[1]}}]+g+g)$ and $\Gamma(Z \to b\bar{b}[{^3P_J^{[1]}}]+b+\bar{b})$. It is worth mentioning that, to satisfy the conservation of $C-$parity, at $B$ factories the process $e^+e^- \to \gamma^{*} \to b\bar{b}[^3P_J^{[1]}]+g+g$ is forbidden. Moreover the center-of-mass energy at $B$ factories (10.6 GeV) is too small to allow for $e^+e^- \to \gamma^{*} \to (b\bar{b})[^3P_J^{[1]}]+b\bar{b}$. From these points of view, for the study of $\chi_b$ the decay of $Z$ boson seems to be more suitable.

For the $^3S_1^{[8]}$ state cases, the NLO QCD corrections can also enhance the LO results significantly, by 10-20 times. The contributions of the $\textrm{NLO}^{*}$ channels are as always sizeable. Similar to $Z \to \chi_c+X$, the CO contributions still account for a large proportion in the total decay width, as listed below
\begin{eqnarray}
\Gamma^{\chi_{b0}}_{\textrm{CO}} / \Gamma^{\chi_{b0}}_{\textrm{\textrm{CS+CO}}}&=&(37.8 - 40.6)\%, \nonumber \\
\Gamma^{\chi_{b1}}_{\textrm{CO}} / \Gamma^{\chi_{b1}}_{\textrm{\textrm{CS+CO}}}&=&(51.5 - 63.3)\%, \nonumber \\
\Gamma^{\chi_{b2}}_{\textrm{CO}} / \Gamma^{\chi_{b2}}_{\textrm{\textrm{CS+CO}}}&=&(83.0 - 83.9)\%.
\end{eqnarray}
$\Gamma^{\chi_{bJ}}_{\textrm{CO}}$ represents the sum of the NLO and $\textrm{NLO}^{*}$ contributions. Regarding the ratios of $\Gamma_{\chi_{b1}}/\Gamma_{\chi_{b0}}$ and $\Gamma_{\chi_{b2}}/\Gamma_{\chi_{b0}}$, the NRQCD predictions are still far different from that built on the CS mechanism,
\begin{eqnarray}
\textrm{CS}&:&~~~\Gamma_{\chi_{b1}} / \Gamma_{\chi_{b0}}= 1.175 - 1.188, \nonumber \\
\textrm{NRQCD}&:&~~~\Gamma_{\chi_{b1}} / \Gamma_{\chi_{b0}} = 1.868 - 1.923, \nonumber \\
\textrm{CS}&:&~~~\Gamma_{\chi_{b2}} / \Gamma_{\chi_{b0}} = 0.626 - 0.657, \nonumber \\
\textrm{NRQCD}&:&~~~\Gamma_{\chi_{b2}} / \Gamma_{\chi_{b0}} = 2.286 - 2.420,
\end{eqnarray}
which can be utilized to check the validity of the CO mechanism.

By adopting the branching ratios of $\chi_b $ to $\Upsilon(1S)$ \cite{Tanabashi:2018oca},
\begin{eqnarray}
\textrm{Br}(\chi_{b0} \to\Upsilon(1S)+\gamma)&=&1.94\%, \nonumber \\
\textrm{Br}(\chi_{b1} \to \Upsilon(1S)+\gamma)&=&35.0\%, \nonumber \\
\textrm{Br}(\chi_{b2} \to \Upsilon(1S)+\gamma)&=&18.8\%,
\end{eqnarray}
we obtain
\begin{eqnarray}
\Gamma(Z \to \Upsilon(1S)+X)|_{\chi_b-\textrm{feeddown}}=(0.15 \sim 0.49) \times 10^{-6}.
\end{eqnarray}

Considering that for $Z \to Q\bar{Q}[^3S_1^{[8]}]+X$ the NLO QCD corrections can enhance the LO results quite significantly, it is interesting and natural to take a brief discussion on the NNLO effect. As stated before, this significant enhancement can be attributed to the kinematic enhancements via the $^3S_1^{[8]}$ single-gluon-fragmentation diagrams. Since the SGF topology has emerged at the NLO level, the NNLO-level diagrams might not enhance the NLO results by orders. Of course whether this is indeed the case depends on the future accomplishment of the NNLO calculations.
\section{Summary}

In this paper, we have systematically investigated the decay of $Z$ boson into $\chi_c$ and $\chi_b$, respectively. We find that the branching ratio for $Z \to \chi_c+X$ is on the order of $10^{-5}$, and $10^{-6}$ for the $\chi_b$ case, which implies that these decay processes are able to be detected. It is observed that, the $^3S_1^{[8]}$ single-gluon-fragmentation diagrams that first emerge at the NLO level can enhance the LO results by about 2-3 orders for $c\bar{c}$, and 10-20 times for $b\bar{b}$. For the same reason, the $\textrm{NLO}^{*}$ processes can also contribute considerably, about $50\%$ of the NLO results. Consequently, the CO contributions will play a vital (even dominant) role in the decay process of $Z \to \chi_c(\chi_b)+X$. Moreover, including the CO channels will thoroughly change the CS predictions on the ratios of $\Gamma(\chi_{c2})/\Gamma(\chi_{c0})$, $\Gamma(\chi_{c1})/\Gamma(\chi_{c0})$, $\Gamma(\chi_{b1})/\Gamma(\chi_{b0})$, and $\Gamma(\chi_{b2})/\Gamma(\chi_{b0})$, which can be regarded as an outstanding probe to distinguish between the CS and CO mechanism. For the CS channels, the heavy quark-antiquark pair associated process, $Z \to Q\bar{Q}[^3P_J^{[1]}]+Q\bar{Q}$, plays a leading role. However, the process of $Z \to Q\bar{Q}[^3P_J^{[1]}]+g+g$ can also provide non-negligible contributions, especially for the $\chi_b$ cases. Taking into account the $\chi_{cJ}$ and $\chi_{bJ}$ feeddown contributions respectively, we find $\Gamma(Z \to J/\psi+X)|_{\chi_c-\textrm{feeddown}}=(0.28 - 2.4) \times 10^{-5}$ and $\Gamma(Z \to \Upsilon(1S)+X)|_{\chi_b-\textrm{feeddown}}=(0.15 - 0.49) \times 10^{-6}$. In summary, the decay of $Z$ boson into $\chi_c(\chi_b)$ is an ideal laboratory to further identify the significance of the color-octet mechanism.

\section{Acknowledgments}
\noindent{\bf Acknowledgments}:
We would like to thank Wen-Long Sang for helpful discussions on the treatments on $\gamma_5$. This work is supported in part by the Natural Science Foundation of China under the Grant No. 11705034., by the Project for Young Talents Growth of Guizhou Provincial Department of Education under Grant No. KY[2017]135., and the Project of GuiZhou Provincial Department of Science and Technology under Grant No. QKHJC[2019]1160.\\


\begin{thebibliography}{1}

\bibitem{Bodwin:1994jh}
  G.~T.~Bodwin, E.~Braaten and G.~P.~Lepage,
  ``Rigorous QCD analysis of inclusive annihilation and production of heavy quarkonium,
  Phys.\ Rev.\ D {\bf 51} (1995) 1125
   Erratum: [Phys.\ Rev.\ D {\bf 55} (1997) 5853]
  doi:10.1103/PhysRevD.55.5853, 10.1103/PhysRevD.51.1125.

\bibitem{Braaten:1994vv}
  E.~Braaten and S.~Fleming,
  Color octet fragmentation and the psi-prime surplus at the Tevatron,
  Phys.\ Rev.\ Lett.\  {\bf 74} (1995) 3327
  doi:10.1103/PhysRevLett.74.3327.

\bibitem{Cho:1995vh}
  P.~L.~Cho and A.~K.~Leibovich,
  Color octet quarkonia production,
  Phys.\ Rev.\ D {\bf 53} (1996) 150
  doi:10.1103/PhysRevD.53.150.

\bibitem{Cho:1995ce}
  P.~L.~Cho and A.~K.~Leibovich,
  Color octet quarkonia production. 2.,
  Phys.\ Rev.\ D {\bf 53} (1996) 6203
  doi:10.1103/PhysRevD.53.6203.

\bibitem{Han:2014jya}
  H.~Han, Y.~Q.~Ma, C.~Meng, H.~S.~Shao and K.~T.~Chao,
  $\eta_c$ production at LHC and indications on the understanding of $J/\psi$ production,
  Phys.\ Rev.\ Lett.\  {\bf 114} (2015) no.9,  092005
  doi:10.1103/PhysRevLett.114.092005.

\bibitem{Zhang:2014ybe}
  H.~F.~Zhang, Z.~Sun, W.~L.~Sang and R.~Li,
  Impact of $\eta_c$ hadroproduction data on charmonium production and polarization within NRQCD framework,
  Phys.\ Rev.\ Lett.\  {\bf 114} (2015) no.9,  092006
  doi:10.1103/PhysRevLett.114.092006.

\bibitem{Gong:2013qka}
  B.~Gong, L.~P.~Wan, J.~X.~Wang and H.~F.~Zhang,
  Complete next-to-leading-order study on the yield and polarization of $\Upsilon(1S,2S,3S)$ at the Tevatron and LHC,
  Phys.\ Rev.\ Lett.\  {\bf 112} (2014) no.3,  032001
  doi:10.1103/PhysRevLett.112.032001.

\bibitem{Feng:2015wka}
  Y.~Feng, B.~Gong, L.~P.~Wan and J.~X.~Wang,
  An updated study of $\Upsilon$ production and polarization at the Tevatron and LHC,
  Chin.\ Phys.\ C {\bf 39} (2015) no.12,  123102
  doi:10.1088/1674-1137/39/12/123102.

\bibitem{Wang:2012is}
  K.~Wang, Y.~Q.~Ma and K.~T.~Chao,
  $\Upsilon(1S)$ prompt production at the Tevatron and LHC in nonrelativistic QCD,
  Phys.\ Rev.\ D {\bf 85} (2012) 114003
  doi:10.1103/PhysRevD.85.114003.

\bibitem{Han:2014kxa}
  H.~Han, Y.~Q.~Ma, C.~Meng, H.~S.~Shao, Y.~J.~Zhang and K.~T.~Chao,
  $\Upsilon(nS)$ and $\chi_b(nP)$ production at hadron colliders in nonrelativistic QCD,
  Phys.\ Rev.\ D {\bf 94} (2016) no.1,  014028
  doi:10.1103/PhysRevD.94.014028.

\bibitem{Butenschoen:2009zy}
  M.~Butenschoen and B.~A.~Kniehl,
  Complete next-to-leading-order corrections to J/psi photoproduction in nonrelativistic quantum chromodynamics,
  Phys.\ Rev.\ Lett.\  {\bf 104} (2010) 072001
  doi:10.1103/PhysRevLett.104.072001.

\bibitem{Sun:2017wxk}
  Z.~Sun and H.~F.~Zhang,
  QCD corrections to the color-singlet $J/\psi$ production in deeply inelastic scattering at HERA,
  Phys.\ Rev.\ D {\bf 96} (2017) no.9,  091502
  doi:10.1103/PhysRevD.96.091502.

\bibitem{Sun:2017nly}
  Z.~Sun and H.~F.~Zhang,
  QCD leading order study of the $J/\psi$ leptoproduction at HERA within the nonrelativistic QCD framework,
  Eur.\ Phys.\ J.\ C {\bf 77} (2017) no.11,  744
  doi:10.1140/epjc/s10052-017-5323-6.

\bibitem{Zhang:2009ym}
  Y.~J.~Zhang, Y.~Q.~Ma, K.~Wang and K.~T.~Chao,
  QCD radiative correction to color-octet $J/\psi$ inclusive production at B Factories,
  Phys.\ Rev.\ D {\bf 81} (2010) 034015
  doi:10.1103/PhysRevD.81.034015.

\bibitem{Butenschoen:2012px}
  M.~Butenschoen and B.~A.~Kniehl,
  J/psi polarization at Tevatron and LHC: Nonrelativistic-QCD factorization at the crossroads,
  Phys.\ Rev.\ Lett.\  {\bf 108} (2012) 172002
  doi:10.1103/PhysRevLett.108.172002.

\bibitem{Chao:2012iv}
  K.~T.~Chao, Y.~Q.~Ma, H.~S.~Shao, K.~Wang and Y.~J.~Zhang,
  $J/\psi$ Polarization at Hadron Colliders in Nonrelativistic QCD,
  Phys.\ Rev.\ Lett.\  {\bf 108} (2012) 242004
  doi:10.1103/PhysRevLett.108.242004.

\bibitem{Gong:2012ug}
  B.~Gong, L.~P.~Wan, J.~X.~Wang and H.~F.~Zhang,
  Polarization for Prompt J/¦× and ¦×(2s) Production at the Tevatron and LHC,
  Phys.\ Rev.\ Lett.\  {\bf 110} (2013) no.4,  042002
  doi:10.1103/PhysRevLett.110.042002.

\bibitem{Chen:2014ahh}
  L.~B.~Chen, J.~Jiang and C.~F.~Qiao,
  NLO QCD Corrections for $\chi_{cJ}$ Inclusive Production at $B$ Factories,''
  Phys.\ Rev.\ D {\bf 91} (2015) no.9,  094031
  doi:10.1103/PhysRevD.91.094031

\bibitem{Braaten:1999qk}
  E.~Braaten, B.~A.~Kniehl and J.~Lee,
  Polarization of prompt $J/\psi$ at the Tevatron,
  Phys.\ Rev.\ D {\bf 62} (2000) 094005
  doi:10.1103/PhysRevD.62.094005.

\bibitem{Sharma:2012dy}
  R.~Sharma and I.~Vitev,
  High transverse momentum quarkonium production and dissociation in heavy ion collisions,
  Phys.\ Rev.\ C {\bf 87} (2013) no.4,  044905
  doi:10.1103/PhysRevC.87.044905.

\bibitem{Shao:2014fca}
  H.~S.~Shao, Y.~Q.~Ma, K.~Wang and K.~T.~Chao,
  Polarizations of $\chi_{c1}$ and $\chi_{c2}$ in prompt production at the LHC,
  Phys.\ Rev.\ Lett.\  {\bf 112} (2014) no.18,  182003
  doi:10.1103/PhysRevLett.112.182003.

\bibitem{Li:2011yc}
  D.~Li, Y.~Q.~Ma and K.~T.~Chao,
  $\chi_{cJ}$ production associated with a $c\bar c$ pair at hadron colliders,
  Phys.\ Rev.\ D {\bf 83} (2011) 114037
  doi:10.1103/PhysRevD.83.114037.

\bibitem{Ma:2010vd}
  Y.~Q.~Ma, K.~Wang and K.~T.~Chao,
  QCD radiative corrections to $\chi_{cJ}$ production at hadron colliders,
  Phys.\ Rev.\ D {\bf 83} (2011) 111503
  doi:10.1103/PhysRevD.83.111503.

\bibitem{Jia:2014jfa}
  H.~F.~Zhang, L.~Yu, S.~X.~Zhang and L.~Jia,
  Global analysis of the experimental data on $\chi_c$ meson hadroproduction,
  Phys.\ Rev.\ D {\bf 93} (2016) no.5,  054033
  Addendum: [Phys.\ Rev.\ D {\bf 93} (2016) no.7,  079901]
  doi:10.1103/PhysRevD.93.054033, 10.1103/PhysRevD.93.079901.

\bibitem{Klasen:2004tz}
  M.~Klasen, B.~A.~Kniehl, L.~N.~Mihaila and M.~Steinhauser,
  $J/\psi$ plus jet associated production in two-photon collisions at next-to-leading order,''
  Nucl.\ Phys.\ B {\bf 713} (2005) 487
  doi:10.1016/j.nuclphysb.2005.02.009

\bibitem{Harris:2001sx}
  B.~W.~Harris and J.~F.~Owens,
  The Two cutoff phase space slicing method,
  Phys.\ Rev.\ D {\bf 65} (2002) 094032
  doi:10.1103/PhysRevD.65.094032.

\bibitem{Korner:1991sx}
  J.~G.~Korner, D.~Kreimer and K.~Schilcher,
  A Practicable gamma(5) scheme in dimensional regularization,
  Z.\ Phys.\ C {\bf 54} (1992) 503.
  doi:10.1007/BF01559471.

\bibitem{Wang:2013ywc}
  X.~P.~Wang and D.~Yang,
  The leading twist light-cone distribution amplitudes for the S-wave and P-wave quarkonia and their applications in single quarkonium exclusive productions,
  JHEP {\bf 1406} (2014) 121
  doi:10.1007/JHEP06(2014)121.

\bibitem{Feng:2017bdu}
 Y.~Feng, Z.~Sun and H.~F.~Zhang,
 Is the color-octet mechanism consistent with the double $J/\psi$ production measurement at B-factories?,
 Eur.\ Phys.\ J.\ C {\bf 77}, 221 (2017).

\bibitem{Sun:2018rgx}
  Z.~Sun, X.~G.~Wu, Y.~Ma and S.~J.~Brodsky,
  Exclusive production of $J/\psi+\eta_c$ at the $B$ factories Belle and Babar using the principle of maximum conformality,
  Phys.\ Rev.\ D {\bf 98} (2018) no.9,  094001
  doi:10.1103/PhysRevD.98.094001

\bibitem{Wang:2004du}
 J.~X.~Wang,
 Progress in FDC project,
 Nucl.\ Instrum.\ Meth.\ A {\bf 534}, 241 (2004).

\bibitem{Tanabashi:2018oca}
  M.~Tanabashi {\it et al.},
  Review of Particle Physics,
  Phys.\ Rev.\ D {\bf 98} (2018) no.3,  030001.
  doi:10.1103/PhysRevD.98.030001.

\bibitem{Acciarri:1998iy}
  M.~Acciarri {\it et al.} [L3 Collaboration],
  Heavy quarkonium production in $Z$ decays,
  Phys.\ Lett.\ B {\bf 453} (1999) 94.
  doi:10.1016/S0370-2693(99)00280-4.



\end{thebibliography}
\end{document}